\documentstyle[twocolumn,aps,prl,epsf] {revtex}
\begin{document}
%
%\preprint{$\begin{array}{l}
%\hbox{SHEP\  96-02} \\
%\hbox{hep-th/9601128}
%\end{array}$}
%
%\vspace{0.25in}
%
%\title{
{\bf On the Fixed-Point Structure of Scalar Fields}

\vspace{0.25in}

In a recent Letter \cite{HH}, certain
properties of the Local Potential
Approximation  (LPA) to the Wilson renormalization group  
were uncovered, which led the authors to conclude that
 $D>2$ dimensional $O(N)$ 
scalar field theories endowed with {\sl non-polynomial} interactions 
allow for a continuum of renormalization group
fixed points (FPs), 
and that around the Gaussian FP, asymptotically
free (actually relevant) interactions exist. If true, this could
herald very important new physics, particularly
for the Higgs sector of the Standard Model ($N=D=4$) \cite{HH}.
Continuing work \cite{HH2} in support of these ideas, has
motivated us to point out that we previously studied the same properties
and showed that they lead to very different conclusions \cite{trunc}. 
Indeed, in as much as the statements in ref.\cite{HH} are correct, they
point to some deep and beautiful facts about the LPA and its 
generalisations \cite{revi}, but however no new physics. 

The LPA, which has a  long history \cite{truncm}, approximates
the renormalization group
 by allowing interactions only in the form of a general 
effective potential $V(\phi,t)$. 
(Here $t=\ln(\Lambda_0/\Lambda)$, where $\Lambda_0/\Lambda$ is the
overall/effective cutoff). 
The sharp cutoff form \cite{nico} can be written as:
\begin{equation}
\hskip-.2cm  {\dot V}+ d\phi V'-DV=
(N-1)\ln(1+\frac{V'}{\phi}) +\ln(1+V'').
\end{equation}
 Here we have written $d=D/2-1$,
 ${\dot V}\equiv\frac{\partial V}{\partial t}$ 
 and ${}^\prime\equiv\frac{\partial}{\partial \phi}$. Ref.\cite{HH} computed
the Taylor expansion coefficients
$u_{2n}$:
$V=\sum_{n=0}^\infty  \zeta^{2-2n} u_{2n} \phi^{2n}$, where 
$\zeta=(4\pi)^{D/4}\sqrt{\Gamma(D/2)}$.
For the FPs {\it i.e.} ${\dot V}=0$,
these coefficients can all be solved in
terms of the mass coefficient $\sigma=u_2>-1$. The fact that $\sigma$
appears otherwise to be arbitrary, led the authors to conclude that 
a continuum of FPs exist. If this were correct it would
be difficult to understand why universality
is seen experimentally (including in simulations)
in the continuous phase transitions of many different systems, 
where it is certainly not the case that the 
(bare) effective potential is always polynomial.  
Indeed if such a continuum really did exist for (1) (with
$D>2$ \cite{revi}) we would have to conclude that the LPA was too severe
an approximation. Fortunately some magic occurs \cite{trunc,revi}: all
but a few FP solutions to (1) are unacceptable since they have singular 
derivatives at some critical
value of the field $\phi=\phi_c$, and fail to exist 
for $\phi>\phi_c$. We plot $\phi_c$ in fig.1, for
$N=D=4$. We see that only the Gaussian FP
potential $V\equiv0$ exists for all $\phi\ge0$, supporting the standard lore
on triviality. Furthermore, 
for values of $\sigma$ around the minimum in fig.1, $\phi_c$ is the singularity
closest to the origin in the $\phi$ complex plane, implying that $u_{2n}\sim
\phi_c(\sigma)^{-2n}$ for large enough $n$. Thus, these $u_{2n}(\sigma)$ 
have a local
maximum at $\sigma=-0.64$.  Halpern {\it et al}
found this maximum and attributed it to a  maximum density of 
FPs, however we see that it actually corresponds to the most singular possible putative FP potential. 
\centerline{
\epsfxsize=.7\hsize\epsfbox{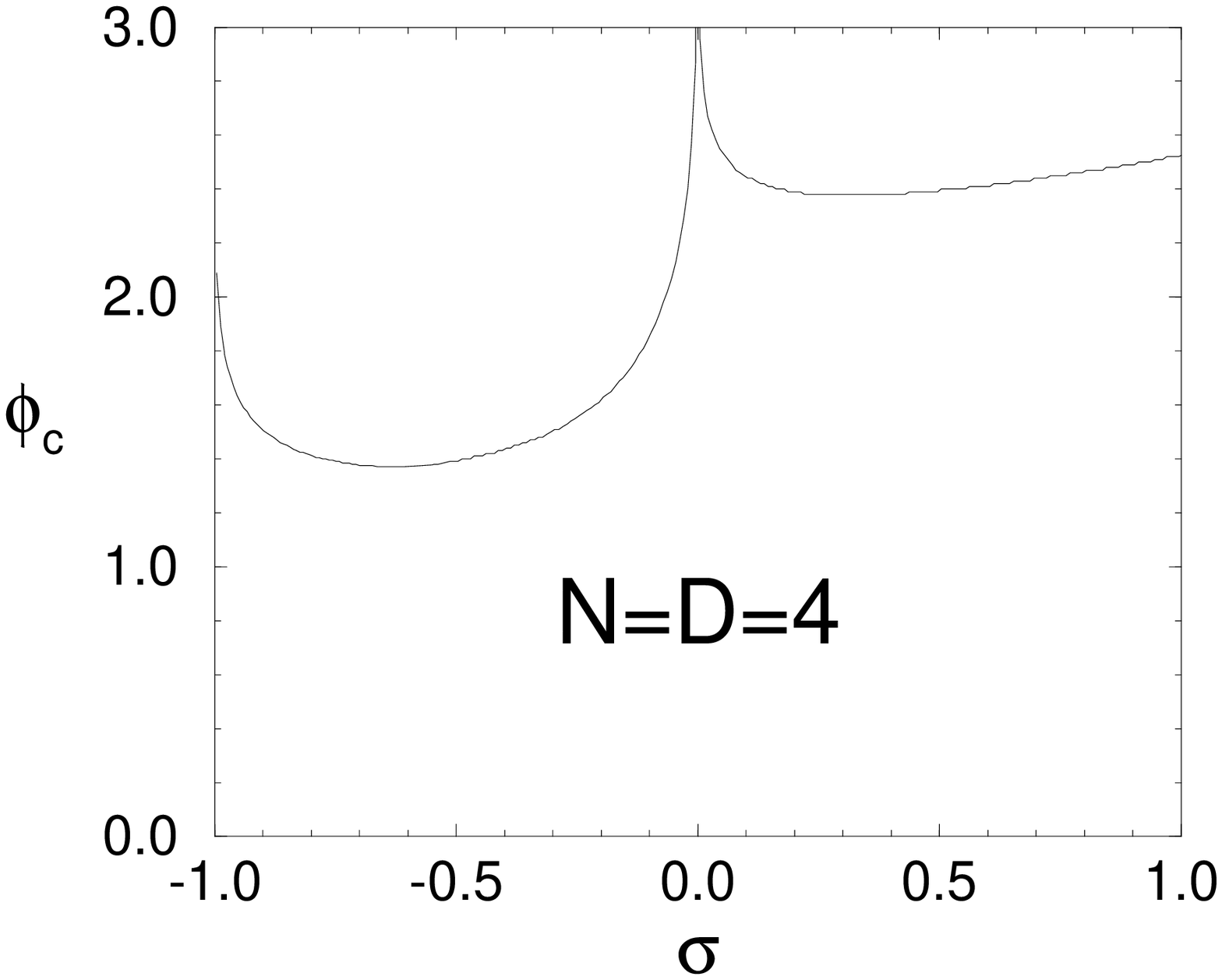}}  
\bigskip
\centerline{\vbox{\noindent {\bf Fig.1.}
Numerical results for $\phi_c(\sigma)$.}}

Around any FP, for finite $\phi$, 
the form of the allowed interactions may be
studied by linearization: $\delta V(\phi,t)=\epsilon v(\phi) e^{\lambda t}$.
From this we tentatively deduce the existence of a small (renormalized)
coupling 
$g(t)=\epsilon e^{\lambda t}$ of (scaling) dimension $\lambda$. However
once again, experimentally we know that $\lambda$ is quantized,
and it is crucial that the LPA reproduce this. Studying the limit 
$\phi\to\infty$, we find that the $v(\phi)$ divide into two
classes: quantized-$\lambda$ perturbations (QPs) that 
behave as $\sim\phi^{(D-\lambda)/d}$, and
non-quantized-$\lambda$ perturbations  (NQPs) that behave as 
$\sim\phi^q\exp(c\phi^p)$ with $c$ and $p$ positive.
It is $\lambda>0$ choices of the latter that
ref.\cite{HH} argues give asymptotically free interactions. We have argued
\cite{trunc} that the NQPs lead to singular
potentials at some $t>0$ \cite{foo}, but irrespective of this, we can see that 
NQPs do not scale as required, since for {\sl any finite}
$\epsilon$ the right hand side of (1) 
contributes negligably as $\phi\to\infty$
and mean field evolution takes over: $\delta V(\phi,t)\sim\epsilon e^{Dt}
v(\phi e^{-dt})$. While this $t$ evolution can still be absorbed into $g(t)$
 in the case of the QPs, it cannot for the NQPs. Indeed
at the Gaussian FP, where the QPs correspond to Laguerre polynomials, 
it follows that as soon as $t>0$ the NQPs become integrable
%evolves so that as soon as $t>0$, it is integrable
with the Laguerre weight and may be reexpanded in terms of the QPs.
Thus, clearly the QPs already span all the continuum physics. 

Further pertinent observations may be found in 
refs.\cite{trunc,revi,truncm}. Our 
conclusions hold for all forms of the LPA.

\medskip\noindent
{\bf Tim R. Morris}\hfill SHEP 96-02\par
Physics Department, University of Southampton,\par
Southampton, SO17 1BJ, U.K.\par


\begin{references}
\bibitem{foo}
They are already singular at $t=0$, when $N=\infty$.
\end{references}
\end{document}